\documentclass[onecolumn, ,amssymb, amsmath, nofootinbib,tightenlines, nobibnotes, aps,prl,epsfig]{revtex4}
 \usepackage{amssymb}
   \usepackage{amsmath}
    \usepackage{amsfonts}
     \usepackage{epsfig}
      \usepackage{bm}
\usepackage{graphicx}
\usepackage{dcolumn}
\usepackage{bm}
\begin{document}
\preprint{APS/123-QED}
\title{Investigation of nonlinear dipole cross sections }

\author{G.R. Boroun}%
 \email{ boroun@razi.ac.ir }
\affiliation{ Department of physics, Razi University, Kermanshah
67149, Iran}

\author{B.Rezaei}%
 \email{ brezaei@razi.ac.ir }
\affiliation{ Department of physics, Razi University, Kermanshah
67149, Iran}

\date{\today}
\begin{abstract}
The nonlinear contributions to the Golec-Biernat-Wusthoff (GBW) and Bartels-Golec-Kowalski (BGK) frameworks, following the approach of Peredo and Hentschinski [M. A. Peredo and M. Hentschinski, Phys. Rev. D \textbf{109}, 014032 (2024)], are examined through their effects on the gluon density and the resulting dipole cross section. In particular, the low-$x$ gluon distribution entering the modified BGK framework is supplemented by nonlinear corrections. This formulation allows the influence of such corrections on the dipole cross section of heavy nuclei to be assessed. At fixed $\sqrt{s}$ and $\mu^2$, the dipole cross section is evaluated down to the kinematic lower limit $x_{\mathrm{min}}=\mu^2/s$ for center-of-mass energies of $\sqrt{s}=89~\mathrm{GeV}$ and $3400~\mathrm{GeV}$. We also investigate the effects of multiple scattering on the coherent photoproduction cross sections of the vector mesons $\Psi(2s)$ and $J/\Psi$ on a lead target.\\ 


\end{abstract}
\maketitle
\subsection{1. Introduction}

 At small Bjorken-$x$, where QCD enters its high-energy regime, the gluon density of the proton
rises rapidly as $x$ decreases. The corresponding growth of sea-quark densities is accompanied by nonlinear contributions associated with gluon recombination,
in which two gluon ladders merge into a gluon or a quark-antiquark pair. Such recombination effects are described by the Gribov-Levin-Ryskin-Mueller-Qiu (GLR-MQ)
framework \cite{GLR1,GLR2,GLR3}, with several phenomenological applications discussed in Refs.~\cite{Sarma1,Sarma2,Boroun1,Boroun2,Boroun3,Guzey1,Guzey2}. These
nonlinear contributions to the small-$x$ gluon density may become important in scenarios involving gluonic hot spots at future facilities, including the Large
 Hadron-Electron Collider (LHeC) \cite{LHeC} and the Future Circular Collider (FCC) \cite{FCC}, where ep DIS can reach values of $x$ of about $10^{-4}$ and $10^{-6}$,
  respectively.\\

The Color Glass Condensate (CGC) \cite{CGC1,CGC2,CGC3,CGC4,CGC5,CGC6,CGC7} provides a more complete description of the gluonic component of the hadron wave
 function and constitutes a semi-classical effective field theory for small-$x$ gluons \cite{Salazer1,Navarra1}. Its Balitsky-Kovchegov (BK) formulation
  \cite{Bali1,Kov1,Kov2}, corresponding in the large-$N_c$ limit to the hierarchy of Wilson-line equations, is formulated naturally in the color dipole picture
  (CDP) \cite{Nikolaev1,Mueller1}. Within the CDP, the $\gamma^{*}p$ interaction proceeds through the fluctuation of the virtual photon into a quark-antiquark pair,
  its interaction with the proton through a color dipole, and the subsequent recombination into the virtual photon \cite{Machado1,Kowalski1}.\\

 We study nonlinear contributions to the dipole description by modifying the corresponding dipole cross sections. For this purpose, the GBW \cite{GBW1}, IIM \cite{IIM1}, and BGK  \cite{BGBK1} dipole parameterizations, together with the GLR-MQ nonlinear evolution framework \cite{GLR1,GLR2,GLR3}, are considered within the color dipole picture. In the GBW formulation, the dipole cross section is written in the eikonal form
 \begin{eqnarray}\label{GBW1_eq}
 \sigma_{\mathrm{dip}}(\widetilde{x},r)=\sigma_{0}\bigg{[}1-\exp{\bigg{(}}-\frac{r^2Q^{2}_{\mathrm{sat}}}{4}\bigg{)}\bigg{]}.
 \end{eqnarray}
 The saturation momentum entering this expression is specified by \begin{eqnarray}\label{saturation_eq}
 Q_{\mathrm{sat}}^{2}(x)=Q_{0}^{2}(\widetilde{x}/x_{0})^{-\lambda},
 \end{eqnarray}
 with $Q_0^2=1~\text{GeV}^2$.  In this form, the dependence of the dipole cross section is expressed through the scaling combination $rQ_{\mathrm{sat}}$. Because the photon
wave function contains the quark-mass dependence of the $q\overline{q}$ state, the Bjorken variable appearing in the dipole cross section is rescaled for
 charm production as \cite{GBW2}
\begin{eqnarray}\label{rescaling_eq}
 x{\rightarrow}\widetilde{x}=x\bigg{(}1+\frac{4m_{c}^{2}}{Q^2}\bigg{)}.
 \end{eqnarray}
 QCD evolution is introduced in the BGK model through the gluon density $xg(x,\mu^2)$ evaluated at the scale $\mu^2=C/r^2+\mu_{0}^{2}$, as the dipole cross
  section is defined as
\begin{eqnarray}\label{BGBK1_eq}
 \sigma_{\mathrm{dip}}(\widetilde{x},r)=\sigma_{0}\bigg{[}1-\exp{\bigg{(}}-\frac{\pi^2r^2\alpha_{s}(\mu^2)xg(\widetilde{x},\mu^2)}{3\sigma_{0}}\bigg{)}\bigg{]}.
 \end{eqnarray}
For small dipole sizes and/or sufficiently small gluon densities, Eq.~(\ref{BGBK1_eq}) approaches its leading-twist limit, namely
 \begin{eqnarray}\label{Collinear_eq}
\sigma_{\mathrm{dip}}(\widetilde{x},r){\simeq}\frac{\pi^2}{3}r^2\alpha_{s}(\mu^2)xg(\widetilde{x},\mu^2).
\end{eqnarray}
For a target with a large gluon density, the dipole cross section can instead be represented as \begin{eqnarray}\label{dipb1_eq}
 \sigma_{\mathrm{dip}}(\widetilde{x},r)=\int d^{2}b\frac{d\sigma_{\mathrm{dip}}}{d^{2}b}. \end{eqnarray}
Here $b$ is the impact parameter of the dipole center measured from the proton center, and the corresponding differential cross section reads
\begin{eqnarray}\label{dipb2_eq}
 \frac{d\sigma^{p}_{\mathrm{dip}}}{d^{2}b}=2\Big{[}1- \exp\Big{(}-\frac{\pi^{2}r^{2}\alpha_{s}(\mu^{2})xg(\widetilde{x},\mu^{2})T(b)}{2N_{c}}\Big{)} \Big{]}.
\end{eqnarray}
The profile function $T(b)$ is fitted to the data in the exponential form \begin{eqnarray}\label{Tb_eq}
T(b)=\frac{1}{2{\pi}B_{G}}\exp(-b^{2}/2B_{G}).
\end{eqnarray}
The parameter $B_G$ is $4.25~\mathrm{GeV}^{-2}$ \cite{Kowalski1}. The expression in Eq.~(\ref{dipb2_eq}) is the Glauber-Mueller dipole cross section
\cite{Mueller2}, which can also be derived from the McLerran-Venugopalan model \cite{CGC1,CGC2}.\\
Besides GBW and BGK, the IIM\footnote{ The IIM form used for small
and large dipoles
 is parameterized by \begin{eqnarray}\label{IIM_eq}
\sigma_{\mathrm{dip}}(\widetilde{x},r)=\sigma_{0}n_{0}\bigg{(}\frac{r^2Q_{\mathrm{sat}}^{2}}{4}\bigg{)}^{\gamma_{\mathrm{sat}}
+\frac{\ln(2/rQ_{\mathrm{sat}})}{k\lambda
Y}}\Theta(r-R_{\mathrm{sat}})
+\sigma_{0}\bigg{[}1-e^{-a\ln^{2}(brQ_{\mathrm{sat}})}\bigg{]}\Theta(R_{\mathrm{sat}}-r),
 \end{eqnarray}
 where $R_{\mathrm{sat}}=2/Q_{\mathrm{sat}}$, $Y=\ln(1/x)$, and the remaining coefficients follow from the continuity requirements at $rQ_{\mathrm{sat}}=2$, as
  given in Refs.~\cite{Navarra1,Machado1}.}  parameterization provides a phenomenological description of the dipole cross section motivated by BFKL dynamics, incorporating features associated with both LO and NLO evolution \cite{Mueller3,Beuf1}. In the CGC
approach, the eikonal representation is written as
 \begin{eqnarray}\label{Tb_eq}
\sigma_{\mathrm{dip}}(\widetilde{x},r)=2{\int}d^{2}\mathbf{b}\mathcal{N}(\widetilde{x},r,\mathbf{b}).
 \end{eqnarray}
 Here $\mathcal{N}$ is the forward dipole-target scattering amplitude at impact parameter $\mathbf{b}$.\\

Phenomenological dipole parameterizations continue to be useful alongside fits based on low-$x$ evolution equations, including BK and BFKL, and calculations
 incorporating fixed-order NLO corrections. The approach used here is nevertheless useful for DIS observables whose description is strongly affected by
  nonperturbative QCD effects, such as momentum-transfer $t$ distributions in diffractive or exclusive reactions.  It also provides a relatively simple framework for studying diffractive and exclusive reactions at HERA and the EIC, as well as ultra-peripheral collisions (UPCs) at the LHC, while allowing systematic improvements in the accuracy of theoretical predictions in view of the increasing precision of experimental data. These features motivate the use of phenomenological dipole models in the present analysis.\\

 We investigate nonlinear effects in DIS observables within dipole factorization framework, focusing on  HERA data and predictions for future colliders. In particular, the prescription
 of Ref.~\cite{Peredo} is used to introduce an additional parameter $k$ that controls the strength of nonlinear contributions in the GBW and BGK dipole cross sections. We further consider the corresponding modifications of the GBW and BGK models and compare their dependence on $k$ through the dipole cross sections and the exclusive photoproduction cross sections of $J/\Psi$ and $\Psi(2s)$.\\

\subsection{2. Modified Dipole Cross Sections}

 Ref.~\cite{Peredo} introduces a parameter $k$ into the GBW and BGK dipole cross sections to interpolate between the linear and nonlinear regimes. This is
 achieved through the replacement of the saturation scale according to $Q^{2}_{\mathrm{sat}}{\rightarrow}k Q^{2}_{\mathrm{sat}}$. The resulting modified GBW
 expression is
 \begin{eqnarray}\label{MGBW_eq}
 \sigma_{\mathrm{dip}}(\widetilde{x},r,k)&=&\frac{\sigma_{0}}{k}\bigg{[}k\frac{r^{2}Q^{2}_{\mathrm{sat}}}{4}-\frac{1}{2}\bigg{(}k^2\frac{r^{2}Q^{2}_{\mathrm{sat}}}{4}
  \bigg{)}^2+...\bigg{]}=\sigma_{0}\frac{r^{2}Q^{2}_{\mathrm{sat}}}{4}\bigg{[}1+\sum_{n=1}^{\infty}\frac{1}{(1+n)!}\bigg{(}-k.\frac{r^{2}Q^{2}_{\mathrm{sat}}}{4}
  \bigg{)}^n\bigg{]}\nonumber\\ &&=\frac{\sigma_{0}}{k}\bigg{[}1-\exp \bigg{(}-k.\frac{r^{2}Q^{2}_{\mathrm{sat}}}{4}\bigg{)}\bigg{]}.
\end{eqnarray}
In the modified GBW expression, the parameter \(k\) controls the strength of the nonlinear contributions associated with multiple scattering. Its limiting cases are
 \begin{itemize}
 \item $k=0$ corresponds to the linear case, for which $\sigma_{\mathrm{dip}}(\widetilde{x},r)=\sigma_{0}\frac{r^{2}Q^{2}_{\mathrm{sat}}}{4}$.
 \item $k=1$ reproduces the HERA fit of the GBW model.
 \item $k>1$ corresponds to an enhancement of nonlinear effects, for which the saturation region extends to $rQ_{\mathrm{sat}}<2$ and the ratio $\sigma/\sigma_{0}<1$.
\end{itemize}
  For a nuclear target, the corresponding color-dipole cross section is obtained by setting $k=A^{1/3}$:
\begin{eqnarray}\label{Nuclei_eq}
 \sigma^{A}_{\mathrm{dip}}(\widetilde{x},r)=A\sigma_{\mathrm{dip}}(\widetilde{x},r,k=A^{1/3}),
 \end{eqnarray}
 For lead this gives $k\simeq5.92$. The increase of $k$ represents the nuclear enhancement of the saturation scale and, consequently, of the gluon density.
 The nuclear version of the GBW cross section is obtained with the transformations $ \sigma_{0}\rightarrow\sigma^{A}_{0}=A^{\frac{2}{3}}\sigma_{0}, $ and
 $ Q^{2}_{\mathrm{sat}}\rightarrow Q^{2}_{\mathrm{sat},A}=A^{\frac{1}{3}}Q^{2}_{\mathrm{sat}}$.\\

Using the same expansion procedure, Ref.~\cite{Peredo} obtains the BGK modification \begin{eqnarray}\label{MBGK_eq}
 \sigma_{\mathrm{dip}}(\widetilde{x},r,k)=\frac{\sigma_{0}}{k}\bigg{[}1-\exp \bigg{(}-k\frac{r^{2}\pi^2\alpha_{s}(\mu^{2})xg(\widetilde{x},\mu^{2})}
 {3\sigma_{0}}\bigg{)}\bigg{]}. \end{eqnarray}
Thus, the BGK modification follows from the same expansion used in Eq.~(\ref{MGBW_eq}) and has the corresponding behavior.\\
 We next modify the BGK framework by introducing nonlinear contributions directly into the gluon distribution. Absorptive, or gluon-recombination, effects
 at low $x$ and low $Q^2$ have been considered in Refs.~\cite{Boroun3,Guzey1,Guzey2,Cai,Pelicer,Boroun4,Boroun5}, with particular emphasis on nonlinear
 corrections to the gluon density through the GLR-MQ evolution equation:
  \begin{eqnarray}\label{GLRMQ_eq}
 \frac{{\partial}xg(x,\mu^2)}{{\partial}{\ln}\mu^2}=\frac{\alpha_{s}(\mu^2)}{2\pi}P_{gg}{\otimes}xg(x,\mu^2) -\frac{9\alpha^{2}_{s}
 (\mu^2)}{2R^2\mu^2}\int_{x}^{1}\frac{dz}{z}[zg(z,\mu^2)]^2,
 \end{eqnarray}
Here $R$ denotes the correlation radius of the interacting partons. The nonlinear term suppresses the growth of the gluon density, leading to a gluon-shadowing effect associated with the onset of saturation.\\
 Ref.~\cite{Thorne} also considers the connection between a gluon density obtained from a dipole model and the conventional gluon distribution of collinear
  factorization. The relation between the dipole cross section and the unintegrated gluon distribution (UGD) is \begin{eqnarray}\label{ThorneUGD_eq}
 \alpha_{s}f_{g}(x,k_{t}^2)=\frac{3\sigma_{0}}{4\pi^2}k_{t}^4(x/x_{0})^{\lambda}e^{-k_{t}^2(x/x_{0})^\lambda}.
\end{eqnarray}
 The above expression describes the small-$x$ data for $F_2$, $F_L$, exclusive vector-meson production, and the diffractive structure function within the
  GBW saturation framework. The integrated gluon distribution follows from the UGD through $ xg(x,\mu^2)=\int_{0}^{\mu^2}\frac{dk_{t}^{2}}{k_{t}^2}f_{g}(x,k_{t}^2). $
   Consequently, the leading-twist connection between the UGD and integrated gluon density takes the form \begin{eqnarray}\label{Thornegluon_eq}
xg({x},\mu^2)&=&\frac{3\sigma_{0}}{4\pi^2\alpha_{s}}\left[-\mu^2
e^{-\mu^2/Q_{\mathrm{sat}}^2}+Q_{\mathrm{sat}}^2(1-e^{-\mu^2/Q_{\mathrm{sat}}^2})
\right]. \end{eqnarray} Using the prescription of
Eq.~(\ref{MGBW_eq}) and the $k$-dependent treatment of
Ref.~\cite{Peredo}, the nonlinearly modified gluon distribution becomes
\begin{eqnarray}\label{Thorne_eq}
xg(\widetilde{x},r,k)&=&\frac{3\sigma_{0}}{4\pi^2\alpha_{s}}\frac{1}{k}\bigg{[}-\mu^2+\frac{1}{k}(\mu^2+kQ^{2}_{\mathrm{sat}})\bigg{(}\frac{\mu^{2}}
{Q^{2}_{\mathrm{sat}}}\bigg{)} \bigg{(}1+\sum_{n=1}^{\infty}\frac{1}{(1+n)!}\bigg{(}-\frac{\mu^{2}}{kQ^{2}_{\mathrm{sat}}}\bigg)^n\bigg{)}\bigg{]}\nonumber\\
 &&=\frac{3\sigma_{0}}{4\pi^2\alpha_{s}}\frac{1}{k}\bigg{[}-\mu^2+(\mu^2+kQ^{2}_{\mathrm{sat}}) \bigg{(}1-e^{-\mu^2/kQ^{2}_{\mathrm{sat}}}\bigg{)}\bigg{]}.
\end{eqnarray}
The corresponding modified BGK cross section is therefore
 \begin{eqnarray}\label{MBGKgluon_eq}
\sigma_{\mathrm{dip}}(\widetilde{x},r,k)=\frac{\sigma_{0}}{k}\bigg{[}1-\exp \bigg{(}-\frac{r^{2}}{4}\bigg{[}-\mu^2+(\mu^2+kQ^{2}_{\mathrm{sat}})
\bigg{(}1-e^{-\mu^2/kQ^{2}_{\mathrm{sat}}}\bigg{)}\bigg{]}\bigg{)}\bigg{]}.
\end{eqnarray}
For a nucleus with mass number $A$, the relation is $ \sigma^{A}_{\mathrm{dip}}(\widetilde{x},r,k=A^{1/3})=A\sigma_{\mathrm{dip}}(\widetilde{x},r,k=A^{1/3}) $.
This result shows that nonlinear contributions become more pronounced for a heavy nuclear target than for a proton. The following section compares the BGK modification without nonlinear corrections to the gluon density, Eq.~(\ref{MBGK_eq}), with the version including those corrections, Eq.~(\ref{MBGKgluon_eq}), for nucleon and nuclear targets.

In the present study, we consider two related phenomenological modifications of the dipole cross section to investigate the combined effects of nonlinear dynamics at small \(x\). The first modification, following Peredo and Hentschinski [36], introduces the parameter \(k\) directly into the eikonal expression of the dipole cross section, thereby modifying the strength of multiple scattering. The second modification incorporates nonlinear corrections into the gluon distribution through the replacement \(Q_{\rm sat}^2\rightarrow kQ_{\rm sat}^2\), as described in Eq. (17). We then combine these two prescriptions by introducing the modified gluon distribution into the modified BGK cross section, leading to Eq. (18). This procedure is intended to investigate the cumulative influence of nonlinear modifications in the gluon density and the dipole scattering amplitude. We note that possible overlaps between these nonlinear contributions cannot be excluded entirely within this phenomenological approach. Therefore, the resulting cross section should be regarded as an effective prescription for assessing the combined impact of these modifications, rather than as a complete separation of independent nonlinear mechanisms.\\

\subsection{3. Results and Conclusion}

 The coefficient functions obtained from the Fit1 analysis of the HERA data using the dipole cross section reported in Ref.~\cite{GBW2} are listed in Table~I.
 The fit parameters $\mu_0^2$ and $C$ in Table~I are taken from the BGK model (see Table~3 of Ref.~\cite{GBW2}), whereas the remaining parameters, $\sigma_0$,
 $\lambda$, and $x_0$, are taken from the GBW model (see Table~1 of Ref.~\cite{GBW2}). It should be noted that the fit parameters used in this study are specific
 to their respective models and are not necessarily directly transferable to other theoretical frameworks. Accordingly, using parameter values obtained from
 different models does not imply that these parameters are interchangeable or that they provide an equivalent description of the underlying physics. Since the
  main objective of the present study is to investigate the qualitative behavior of different models and compare their predictions, rather than to perform a
  detailed quantitative comparison with experimental data, we retain the original parameter values associated with each model for this purpose.\\

Figure~1 shows the ratio $\sigma/\sigma_{0}$ as a function of the dipole size $r$ for the modified GBW and BGK expressions with $k=1$ and $10$. Increasing $k$
from the linear case ($k=0$) to larger values ($k>1$) leads to a reduction in both $\sigma/\sigma_{0}$ and the extent of the saturation region in the dipole-size
range. As shown in Fig.~1, the modified GBW and BGK models without nonlinear corrections to the gluon density, given by Eqs.~(\ref{MGBW_eq}) and (\ref{MBGK_eq}),
respectively, exhibit similar behavior for both values of the parameter $k$ ($k=1$ and $10$). The nonlinear effects associated with the $k$-dependent modification
become more pronounced as $k$ increases. In particular, the transition region, characterized by a decrease in $\sigma/\sigma_0$ from 1 to 0.1 at large $r$, changes
as $k$ increases. Accordingly, the saturation region shifts from $r>1$ to approximately $r>0.3$.

In the following, we incorporate the nonlinear behavior of the gluon density into the modified BGK model according to Eq.~(\ref{MBGKgluon_eq}). As shown in Fig.~1,
 the nonlinear corrections to the gluon density become more significant for larger values of $k$ (i.e., $k>1$). The deviation of $\sigma/\sigma_0$ from the result
  for $k=1$ indicates the increasing importance of nonlinear effects in the gluon density, as discussed in Ref.~\cite{KLN}. For $k=1$, these effects are negligible
   at small $r$, whereas they become appreciable mainly for $k>1$ and small dipole sizes ($r<1$). These results provide an estimate of the nonlinear effects
   associated with the gluon density in $eA$ collisions at small $x$ within the color dipole approach. These results provide a phenomenological framework for
   investigating nonlinear gluon dynamics and saturation effects in electron-ion collisions, particularly in the small-$x$ region accessible to future facilities
   such as the Electron-Ion Collider (EIC) \cite{EIC} and the Large Hadron-Electron Collider (LHeC) \cite{LHeC}.\\
\begin{figure}
\centerline{ \includegraphics[width=0.58\textwidth]{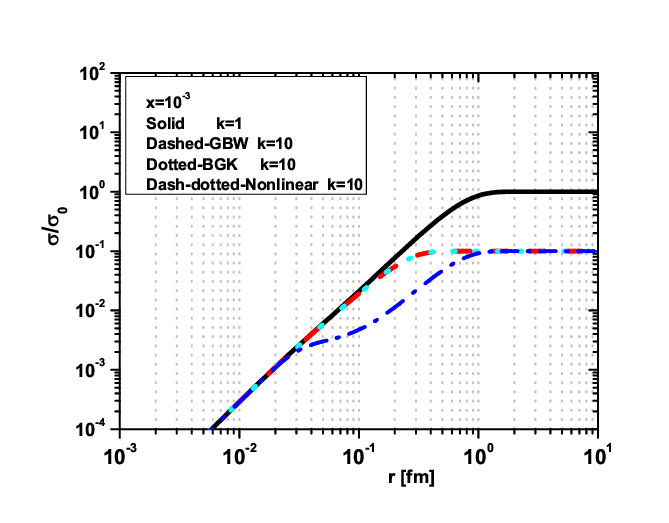}}
 \caption{Ratio $\sigma/\sigma_{0}$ obtained from the modified GBW model (Eq.~(\ref{MGBW_eq})), the modified BGK model (Eq.~(\ref{MBGK_eq})),
  and the modified BGK model including nonlinear corrections to the gluon density (Eq.~(\ref{MBGKgluon_eq})) as a function of the transverse dipole
  size $r$ over a wide range at $x=10^{-3}$. The solid black curve correspond to $k=1$ for the modified GBW and BGK models, while the red dashed and
  cyan dotted curves correspond to $k=10$ for these models. The blue dash-dotted curve represents the modified BGK model with nonlinear corrections to
   the gluon density for $k=10$.} \label{Fig1}
\end{figure}

In Fig.~2, we present $\sigma/\sigma_0$ divided by $A$ for the
light and heavy nuclei $^{12}\mathrm{C}$ and
$^{208}\mathrm{Pb}$, corresponding to $k\simeq
2.289$ and $k\simeq 5.925$, respectively, as a function of $r$ at
$x=10^{-3}$ (left panel) and $x=10^{-6}$ (right panel). Following
Eq.~(\ref{MBGKgluon_eq}), we replace $k$ by $A^{1/3}$ and obtain
\begin{eqnarray}\label{gluonA_eq}
\frac{1}{A}\sigma_{\mathrm{dip}}^{A}(\widetilde{x},r,A)=\frac{\sigma_{0}}{A^{1/3}}\bigg[1-\exp
\bigg(-\frac{r^{2}}{4}\bigg[-\mu^2+(\mu^2+A^{1/3}Q^{2}_{\mathrm{sat}})
\bigg(1-e^{-\mu^2/A^{1/3}Q^{2}_{\mathrm{sat}}}
\bigg)\bigg]\bigg)\bigg].
\end{eqnarray}

The nonlinear effects associated with the gluon density become increasingly significant for larger values of $k$ and smaller $x$, reflecting the
enhanced gluon density in larger nuclei. For the carbon nucleus, nonlinear effects are already visible at the lowest considered value, $x=10^{-6}$,
in the range $0.03~\mathrm{fm}<r<1~\mathrm{fm}$. For the lead nucleus, these effects appear at smaller dipole sizes, approximately
$0.02~\mathrm{fm}<r<1~\mathrm{fm}$, accompanied by a more pronounced depletion in the ratio.\\

\begin{figure}
\centerline{
\includegraphics[width=0.60\textwidth]{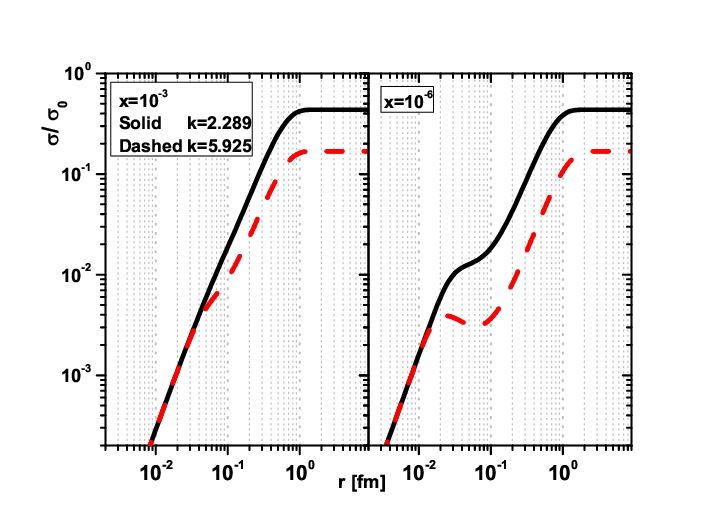}}
\caption{Nonlinear effects associated with the gluon density in
the modified BGK model for $k\simeq2.289$ (solid-black) and
$k\simeq5.925$ (dashed-red) at $x=10^{-3}$ (left panel) and
$x=10^{-6}$ (right panel), shown over a wide range of the dipole
size $r$.}\label{Fig2}
\end{figure}

\begin{figure}
\centerline{
\includegraphics[width=0.60\textwidth]{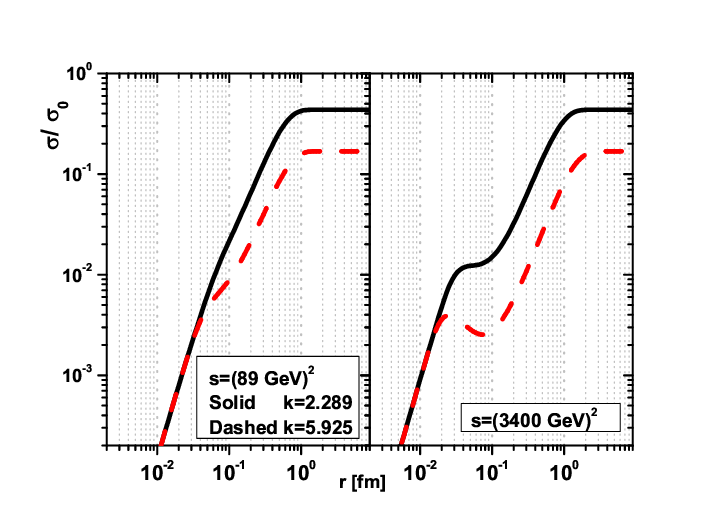}}
\caption{Nonlinear effects associated with the gluon density in
the modified BGK model for $k\simeq2.289$ (solid-black) and
$k\simeq5.925$ (dashed-red) at the center-of-mass energies
$\sqrt{s}=89~\mathrm{GeV}$ (left panel) and
$\sqrt{s}=3.4~\mathrm{TeV}$ (right panel), in the limit
$x_{\mathrm{min}}=Q^2/s$, shown over a wide range of the dipole
size $r$.}\label{Fig3}
\end{figure}

It is worth noting that, at fixed center-of-mass energy
$\sqrt{s}$, where $s$ denotes the squared center-of-mass energy of
the electron-proton or electron-nucleus system, with $s=4E_eE_p$,
and fixed $\mu^2$, the dipole cross section reaches the minimum
value of $x$ given by
 $x_{\mathrm{min}}=\mu^2/s$. This kinematic limit provides a bound on the behavior of the dipole cross section.
 In the high-inelasticity limit,\footnote{The longitudinal structure function in this limit has recently been studied within the color
  dipole picture in Ref.~\cite{Taylor}.},
 the dipole cross section approaches a limiting behavior at $x_{\mathrm{min}}$, which depends on the center-of-mass energy. The leptonic contribution to the DIS cross section has been factorized out, leaving the dependence on the inelasticity $y=1$ in the corresponding reduced cross section. At this limit,
  the modified Bjorken variable is defined as
$\widetilde{x}_{\mathrm{min}}=\mu^2/s+4m_c^2/s$, and the nuclear dipole cross section divided by $A$ can therefore be expressed as a function of the center-of-mass energy squared, $s$, as
\begin{eqnarray}\label{MBGKgluonA_eq}
\frac{1}{A}\sigma^{A}_{\mathrm{dip}}(s,r,A)&=&\frac{\sigma_{0}}{A^{1/3}}\bigg[1-\exp
\bigg(-\frac{r^{2}}{4}\bigg[-\mu^2+\bigg(\mu^2+A^{\frac{1}{3}}\bigg(\frac{\mu^2}{sx_{0}}+
\frac{4m_{c}^2}{sx_{0}} \bigg)^{-\lambda}\bigg)\nonumber\\
&&{\times}\bigg(1-e^{-\mu^2/A^{\frac{1}{3}}\bigg(\frac{\mu^2}{sx_{0}}
+\frac{4m_{c}^2}{sx_{0}} \bigg)^{-\lambda}}
\bigg)\bigg]\bigg)\bigg].
\end{eqnarray}
The corresponding result for $\sigma^{A}_\mathrm{dip}/(A\sigma_0)$ is shown in Fig.~3 for the center-of-mass energies $\sqrt{s}=89~\mathrm{GeV}$ and
$3.4~\mathrm{TeV}$, for the light and heavy nuclei $^{12}\mathrm{C}$ and
$^{208}\mathrm{Pb}$,
 with $k\simeq2.289$ and $k\simeq5.925$, respectively. The ratio is evaluated as a function of $r$ in the limit $\widetilde{x}_{\mathrm{min}}$, corresponding to the
 high-inelasticity regime. The nonlinear effects associated  with the gluon density become more pronounced at high inelasticity and larger values of $k$ (i.e., $k\simeq5.925$),
particularly in the range $0.05~\mathrm{fm}\lesssim r \lesssim1~\mathrm{fm}$ at $\sqrt=89~\mathrm{GeV}$, extending to the range $0.02~\mathrm{fm}\lesssim r\lesssim2~\mathrm{fm}$ at $\sqrt{s}=3.4~\mathrm{TeV}$. Indeed, the nonlinear behavior of the gluon density in the modified BGK model becomes more pronounced with increasing center-of-mass energy.\\

\begin{table}
\centering \caption{Coefficient values obtained from Ref.~\cite{GBW2}.}\label{table:table1}
\begin{minipage}{\linewidth}
\renewcommand{\thefootnote}{\thempfootnote}
\centering
\begin{tabular}{|l|c|c|c|c|c|c|} \hline\noalign{\smallskip}
fit & $m_{c}[\mathrm{GeV}]$ & $\mu^2_{0}[\mathrm{GeV}^2]$ & $C$ & $\sigma_{0}[\mathrm{mb}]$ & $\lambda$ & $x_{0}/10^{-4}$  \\
\hline\noalign{\smallskip}
1 & 1.4 & 1.85$\pm$0.20 & 0.29$\pm$0.05 & $27.32{\pm}0.35$ & $0.248{\pm}0.002$ & $0.42{\pm}0.04 $  \\
\hline\noalign{\smallskip}
\end{tabular}
\end{minipage}
\end{table}

The coherent photoproduction of heavy vector mesons, \(J/\Psi\) and \(\Psi(2s)\), provides an additional probe of the gluon distribution and nonlinear effects at small \(x\) \cite{Peredo}. The gluon distribution entering this expression is obtained from the modified BGK framework. Thus, the collinear expression\footnote{It should be emphasized that the use of the collinear expression in Eq.~(\ref{Cross_eq}) does not replace the dipole framework employed in the preceding sections. Rather, it provides an alternative formulation for calculating the forward vector-meson photoproduction cross section, with the nonlinear effects inherited from the modified BGK gluon distribution.} is used to calculate the forward cross section, while the nonlinear dependence is incorporated through the modified gluon distribution. The process is described by
\begin{eqnarray}\label{Photo_eq}
\gamma(q)+p(p){\rightarrow}V(q')+p(p'),~~~~~V=J/\Psi, \Psi(2s),
\end{eqnarray}
where the cross section is related to the square of the gluon distribution evaluated at the factorization scale \(\mu^2\). The gluon distribution entering this expression is obtained from the modified BGK framework. Thus, the collinear expression is used to calculate the forward cross section, while the nonlinear dependence is incorporated through the modified gluon distribution.
 At leading order in QCD, the differential cross section at zero momentum transfer is given by \cite{Cross1, Cross2}
\begin{eqnarray}\label{Cross_eq}
\frac{d\sigma}{dt}(\gamma{A}{\rightarrow}VA)|_{t=0}=\frac{\Gamma^{V}_{ee}M^{3}_{V}\pi^3}{48\alpha_{\mathrm{em}}}
\bigg[\frac{\alpha_{s}(\mu^2)}{m_{c}^{4}}xg^{A}(x,\mu^2,k)\bigg]^2,
\end{eqnarray}
where $t=(q-q')^2$, and the forward limit $t=0$ is related to the inclusive gluon distribution. The relevant parameters are
$\Gamma^{J/\Psi}_{ee}=5.55{\times}10^{-6}~\mathrm{GeV}$ and
$M_{J/\Psi}=3.1~\mathrm{GeV}$ for the $J/\Psi$ state, and
$\Gamma^{\Psi(2s)}_{ee}=2.33{\times}10^{-6}~\mathrm{GeV}$ and
$M_{\Psi(2s)}=3.770~\mathrm{GeV}$ for the $\Psi(2s)$ state.

In our numerical calculations, the gluon distribution \(xg^A(x,\mu^2,k)\) is evaluated using the modified gluon-density prescription introduced in Eq.~(\ref{Thorne_eq}). For nuclear targets, the parameter \(k=A^{1/3}\) is adopted to account phenomenologically for the enhanced nonlinear effects associated with the nuclear gluon density. This modified distribution is then substituted into Eq.~(\ref{Cross_eq}) to calculate the forward photoproduction cross section. Therefore, the multiple-scattering effects shown in Fig. 4 are incorporated through the nonlinear modification of the gluon distribution, rather than through an explicit eikonal treatment of the vector-meson production amplitude.\\
In Fig.~4, we investigate the effect of multiple scattering on the forward photoproduction cross section, $d\sigma/dt(\gamma A\rightarrow VA)|_{t=0}$, for the coherent production of the $J/\Psi(1s)$ and $\Psi(2s)$ vector mesons on free-proton ($A=1$) and lead ($A=208$) targets at $x=10^{-3}$. The factorization scale $\mu^2$ is treated as a variable in the range $1\leq\mu^2\leq5~\mathrm{GeV}^2$ to examine the scale dependence of the cross section. In addition, the characteristic scales $\mu^2=M_V^2/4$, commonly adopted in leading-order calculations of heavy vector-meson photoproduction, are indicated for each meson. These correspond to approximately $2.403~\mathrm{GeV}^2$ for $J/\Psi$ and $3.553~\mathrm{GeV}^2$ for $\Psi(2s)$. As discussed in Ref.~\cite{Nemchik}, transverse momentum-transfer distributions $d\sigma/dt$ in coherent charmonium electroproduction off nuclei are sensitive to multiple-scattering effects. In the present calculation, these effects are incorporated through the nonlinear modification of the gluon distribution within the modified BGK framework. As shown in Fig.~4, they lead to a slower increase of the forward photoproduction cross section at small values of the factorization scale, particularly for the lead target.\\

\begin{figure}
\centerline{
\includegraphics[width=0.60\textwidth]{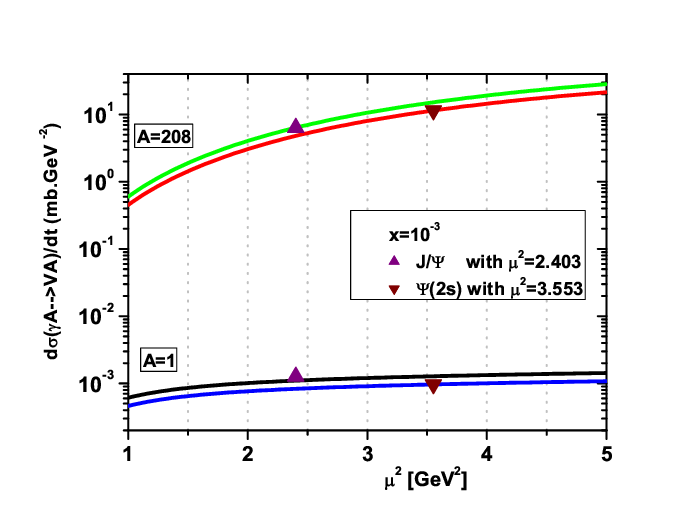}}
\caption{Forward coherent photoproduction cross sections $d\sigma/dt(\gamma{A}{\rightarrow}VA)|_{t=0}$ for $J/\Psi$ and $\Psi(2s)$
 on proton ($k=1$) and lead ($k=A^{1/3}\simeq5.925$) targets at $x=10^{-3}$, as functions of the factorization scale $\mu^2$.
 The solid and dashed curves correspond to $J/\Psi$ and $\Psi(2s)$, respectively. The characteristic cross-section values (purple-up triangle and brown-down triangle) are
 indicated at $\mu^2=M_V^2/4$ for each vector meson.}\label{Fig6}
\end{figure}

In conclusion, we have presented a method based on the dipole cross sections of the modified GBW and BGK models to investigate the relevance of nonlinear corrections to the gluon density. We have incorporated nonlinear corrections into the gluon density of the BGK model and shown that these corrections modify the behavior of the resulting dipole cross section compared with the original BGK model. The approach is based on a parametrization of the gluon density in the low-$x$ region for both nucleons and nuclei. The nonlinear effects become more pronounced for large values of $k$ ($k>1$) and smaller values of $x$, with the relevant dipole-size range depending on the nuclear target and kinematic conditions. The modified BGK model therefore provides a useful framework for estimating the effects of low-$x$ nonlinear corrections to the gluon density and their impact on the dipole cross section in heavy-ion and electron-ion collisions.

We have also investigated the effects of multiple scattering on the differential cross section $d\sigma/dt(\gamma A\rightarrow VA)|_{t=0}$ in the coherent photoproduction of the heavy charmonium states $J/\Psi$ and $\Psi(2s)$ on a lead nucleus within the modified BGK framework. The results indicate that multiple scattering modifies the behavior of $d\sigma/dt$ at small $x$ and low values of the factorization scale $\mu^2$. These results indicate the sensitivity of coherent charmonium photoproduction to nonlinear effects in the gluon density. They may provide a useful basis for further investigations of such effects under different kinematic conditions, including future electron-ion collision measurements at the EIC. Further studies incorporating the specific kinematics of ultraperipheral heavy-ion collisions at the LHC would also be of interest.\\


\subsection{ACKNOWLEDGMENTS}
The authors are grateful to Razi University for the financial
support provided for this project.

\section{DECLARATION OF AI-ASSISTED LANGUAGE EDITING}

The authors have used AI-assisted tools solely for language
editing and improving the readability of the manuscript. The
authors have carefully reviewed and revised the manuscript and
takes full responsibility for
its scientific content.\\

\end{document}